\documentclass[dvipdfmx]{pasj01}

\draft
\usepackage{comment}
\usepackage{color}
\usepackage{lineno}
\usepackage{graphicx}

\begin{document} 
\Received{}
\Accepted{}

\title{X-ray spectroscopy of the accretion disk corona source 2S\,0921$-$630 with Suzaku archival data}

\author{Tomokage \textsc{yoneyama}\altaffilmark{1}%
}
\author{Tadayasu \textsc{dotani}\altaffilmark{2,3}}

\altaffiltext{1}{Institute of Space and Astronautical Science, Japan Aerospace Exploration Agency, 3-1-1 Yoshinodai, Chuo-ku, Sagamihara 252-5210, Japan}

\altaffiltext{2}{Department of Physics, Tokyo Institute of Technology 2-12-1 Ookayama, Meguro-ku, Tokyo 152-8550, Japan}

\altaffiltext{3}{Department of Space and Astronautical Science, School of Physical Science, The Graduate University for Advanced Studies, SOKENDAI, 3-1-1 Yoshinodai, Chuo-ku, Sagamihara 252-5210, Japan}

\email{yoneyama.tomokage@jaxa.jp}


\KeyWords{}

\maketitle

\begin{abstract}
2S\,0921$-$630 is an eclipsing low-mass X-ray binary (LMXB) with an orbital period of $\sim$ 9\,days. Past X-ray observations have revealed that 2S\,0921$-$630 has an extended accretion disk corona (ADC), from which most of the X-rays from the system are emitted. We report the result of our Suzaku archival data analysis of 2S 0921$-$630. The average X-ray spectrum is reproduced with a blackbody emission ($kT_{\rm bb} \sim 0.3$\,keV) Comptonized by optically-thick gas (``Compton cloud''; optical depth $\tau \sim 21$) with a temperature of $\sim 2$\,keV, combined with thirteen emission lines. We find that most of the emission lines correspond to highly ionized atoms: O, Ne, Mg, Si, S, Ar, and Fe. A K$\alpha$ emission line and an absorption edge of semi-neutral iron (Fe I -- XVII) are also detected. The semi-neutral iron K$\alpha$ line is significantly broad with a width of $0.11 \pm 0.02$\,keV in sigma, which corresponds to the Doppler broadening by the Kepler motion at a radius of $\sim 10^9$\,cm. We suggest that the observed semi-neutral iron line originates at the inner part of the accretion disk in the immediate outside of the Compton cloud, i.e., the Compton cloud may have a radius of $\sim 10^9$\,cm. 
\end{abstract}


\section{Introduction}
Low-mass X-ray binaries (LMXBs) have been very important targets in astrophysics since the dawn of the X-ray astronomy, because they produce fundamental information of compact objects and evolution of bynary systems. In spite of the long history of studies of the species, the nature of the X-ray emitting regions in LMXBs is not fully understood, yet. In particular, the geometry of the emitting region in a LMXB of which the compact star is a neutron star is still highly uncertain; it seems there is a substantial amount of diversity in the geometry from LMXB to LMXB. Among common directly-observable parameters of LMXBs, the orbital period is a key parameter, for it is directly related to the size of the Roche lobe and thus that of the accretion disk, the vicinity of which is believed to be a major X-ray emitting region in a LMXB. Accretion disks around black holes and neutron stars emit anisotropic radiation, and accordingly, orbital inclinations also cruicially affect observed properties. Highly inclined systems often show an eclipse for a duration in each orbit. Conversely, observations of eclipsing binaries allow us to constrain their orbital inclinations. In such a system, X-ray emission from the compact star and inner part of its accretion disk can be hidden by the outer part of the accretion disk. Observation of such systems usually show some amount of dip in their light curves during the periods of their eclipses, rather than the total blockage of the emission. Then, most of the X-ray emission from such systems is considered to originate in extended, optically-thin, highly-ionized gas around the accretion disk, called accretion disk corona (ADC: e.g. \cite{white81} for 4U\,1822$-$37, and  \cite{mcclintock82} for 4U\,2129$+$47).

Generally, in an ADC source with a high inclination, emission from its compact star and/or inner part of its accretion disk is shadowed by the outer part of the accretion disk. We thus observe a scattered  X-ray by the extended ADC. ADCe are considered to be completely ionized such that X-ray emission is not attenuated, but only scattered in ADCe. Hence, observed X-ray spectrum scattered by the ADC should not be transformed by absorption and/or stimulated emission. By contrast, the X-ray flux of the primally emitting region is hard to determine, because the observed flux is only a part of scattered X-ray.

The X-ray source 2S\,0921$-$630 is an eclipsing LMXB discovered by \citet{li78} with SAS-3 and is a well-known ADC source (\cite{mason87}). Its dynamical properties have been determined from optical observations: the distance $D \sim 7$\,kpc (\cite{cowley82}), orbital period $P = 9.0026 \pm 0.0001$\,day, orbital inclination $i = 82^\circ.2 \pm 1^\circ.0$ (\cite{ashcraft12}), compact-star mass $M_{\rm 1} = 1.44 \pm 0.10$\,M$_\odot$, and companion-star mass $M_{\rm 2} = 0.35 \pm 0.03$\,M$_\odot$ (\cite{steeghs07}). The mass of 1.44\,M$_\odot$ implies that the compact star is a neutron star, although a possibility of a low-mass black hole has been suggested (\cite{jonker05}; \cite{shahbaz04}) before the accurate mass was determined. The orbital period of 9\,days is one of the longest among LMXBs (\cite{liu07}). \citet{mason87} reported slight softening in the X-ray spectrum during an eclipse in past observations. Although the eclipse is apparent in optical observations, no drastic variation is seen in X-ray observations. It is interpreted that the ADC is so large that shielding by the companion star during the eclipse is ineffective.

High-resolution X-ray spectroscopic studies of the source were reported by \citet{kallman03}. They analyzed the $ASCA$, $Chandra$, and $XMM$-$Newton$ data and found emission lines from highly-ionized ions: O, Ne, Mg, Si, S, and Fe. In their $Chandra$/HETG and $XMM$/EPIC-pn observations, the K$\alpha$ emission line from semi-neutral iron (Fe\,I -- XVII) was also detected. They showed that the semi-neutral K$\alpha$ line is significantly broad with a width of $0.29 \pm 0.15$\,keV in the EPIC-pn spectrum. They suggested that the iron-line broadening originated from a Keplerian broadening at a disk radius of $\sim 10^8$\,cm or a mixture of the fluorescent lines from various ionization states of iron. They reported that the continuum is reproduced with a very hard power law (photon index $\Gamma = 1.1$), which implies that the primally X-ray, such as the thermal emission from the inner part of the accretion disk, is reprocessed with inverse Compton scattering by a hot plasma (``Compton cloud''), while the properties of the primally X-ray source and the Compton cloud was not revealed in their work.

In this paper, we report ther results of our analysis of $Suzaku$ archival data of 2S\,0921$-$630. In section 2, we describe the observations, data reduction, analysis and results. In section 3, we compare our results with the pervious work by \citet{kallman03} and discuss the properties of the Compton cloud and origin of the semi-neutral iron line.

Errors in tables are in the 90\% confidence level, whereas those in spectral plots and light curves are in $1\sigma$ unless otherwise specified.

\section{Data analysis and results}

\subsection{Data reduction}

We analyze the three sets of $Suzaku$ archive data of 2S\,0921$-$630 obtained in 2007 as listed in table \ref{tab:obs}. We use cleaned event files, to which the standard screening criteria and data corrections are applied. Spectra and light curves for the source and background are extracted from each X-ray Imaging Spectrometer (XIS), XIS0, 1, and 3 with XSELECT V2.4m where the respective regions for extraction are a circle of a $3^\prime$ radius and an annulus of $5^\prime$ inner and $7^\prime$ outer radii. The redistribution matrix files (RMFs) and auxiliary response files (ARFs) are generated with xisrmfgen and xissimarfgen, respectively. For spectral fittings in this work, XSPEC ver12.12.0 is used. In the following analyses, spectra extracted from the front-illuminated (FI) CCDs, XIS0 and 3, are merged to obtain higher statistics. The resultant XIS0$+$3 data are hereafter called the ``XISFI''. For the XISFI, the detector response is generated by averaging the XIS0 and XIS3 responses, i.e., those convoluted from the RMFs and ARFs.

\subsection{Average spectrum}

\begin{figure}[htbp]
 \begin{center}
  \includegraphics[width=8cm]{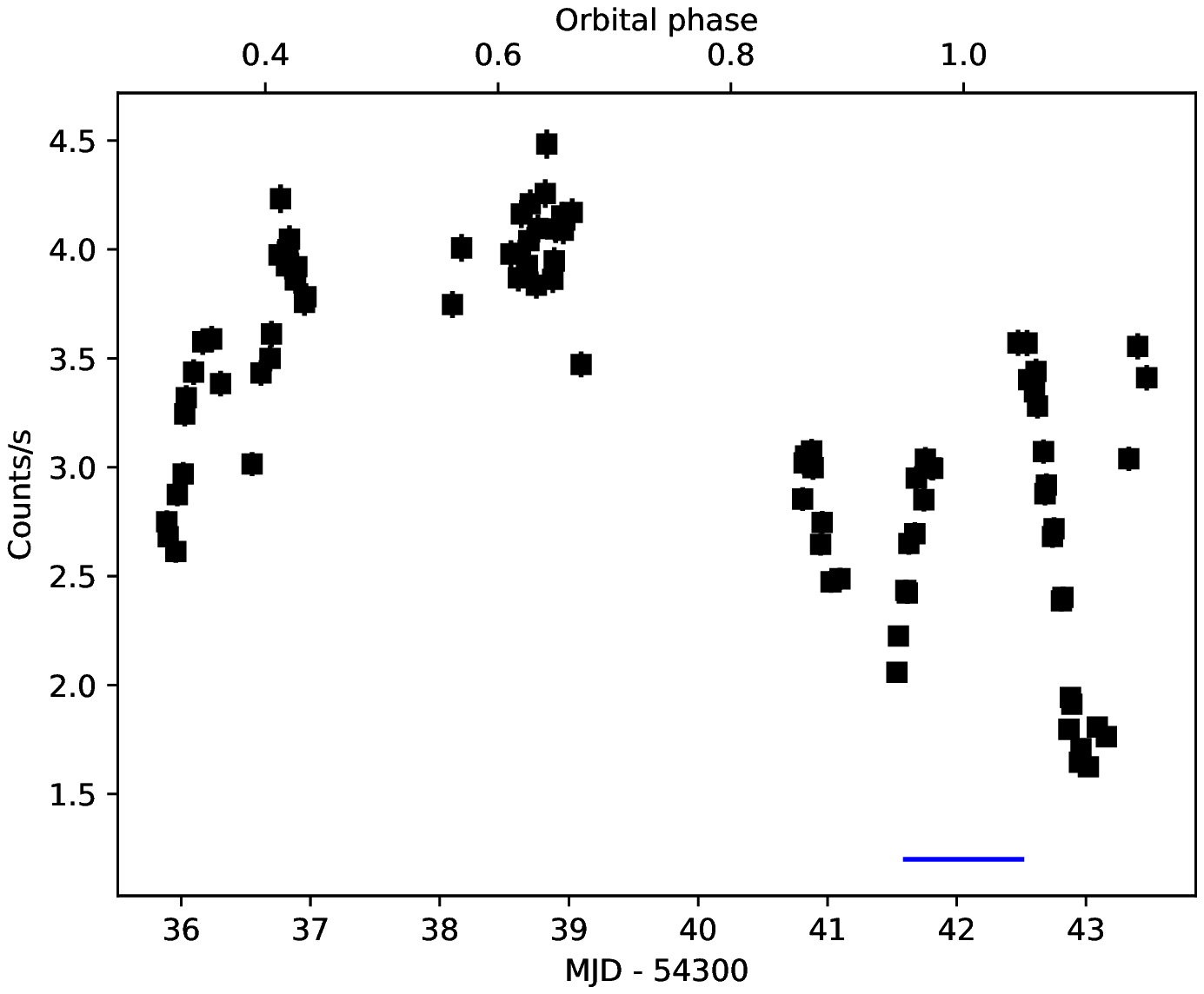} 
 \end{center}
\caption{The 0.4 -- 10.0\,keV light curve of 2S\,0921$-$630 with a time bin of 1000\,s. Only the XIS0 are shown for clarity. Blue horizontal line indicates an eclipse phase.
}\label{fig:lc}
\end{figure}

Figure \ref{fig:lc} shows the light curves of 2S\,0921$-$630 obtained with XIS0 with a time bin of 1000\,s for the  0.4 - 10.0\,keV band. An eclipsing phase at around MJD 54342 derived from an ephemeris produced by \citet{ashcraft12}, which is indicated in the figure, was partially covered in the $Suzaku$ observations. No apparent change in the X-ray flux was observed during the eclipse period. The light curve was dominated by random variations thoughout the observation periods. We thus do not discriminate the data during the eclipse. Though the flux varied significantly during the obsercations, we first analyze the average spectrum to take an advantage of the highest statistics. 

\subsubsection{Spectral analysis}

\begin{figure}[htbp]
 \begin{center}
  \includegraphics[width=8cm]{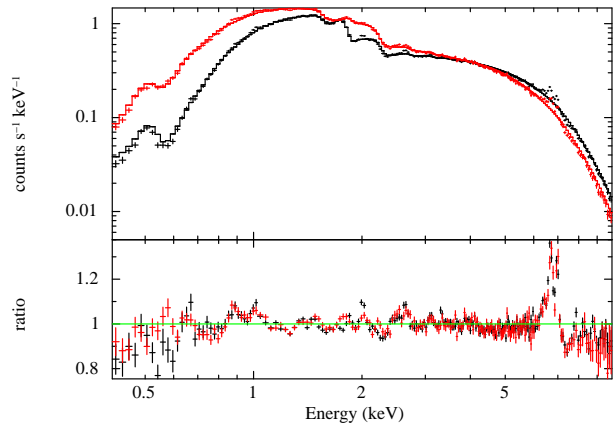} 
 \end{center}
\caption{XISFI ($black$) and XIS1 ($red$) spectra fitted with a blackbody model Comptonized by thermal electrons, plotted together with the ratio of the data to the best-fit model ($bottom$ $pannel$). }\label{fig:xifi1_total_thcbb}
\end{figure}

\begin{figure}[htbp]
 \begin{center}
  \includegraphics[width=8cm]{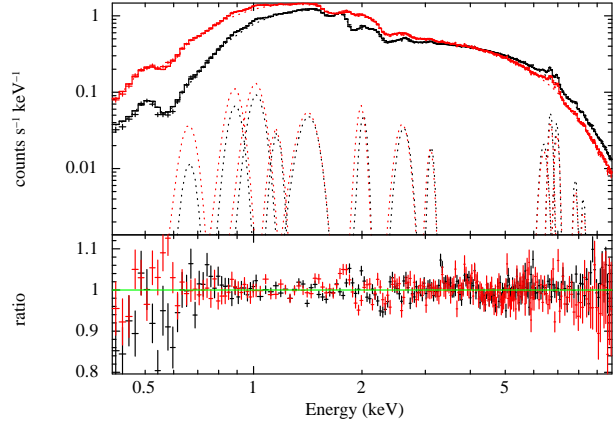} 
 \end{center}
\caption{XISFI ($black$) and XIS1 ($red$) spectra fitted with a blackbody model Comptonized by thermal electrons, 13 Gaussians and an absorption edge plotted together with the ratio of the data to the best-fit model ($bottom$ $pannel$). The dashed lines correspond to individual model components.}\label{fig:xifi1_total_thcbb11g1e}
\end{figure}

We perform simultaneous $\chi^2$-fitting for the XISFI and XIS1 spectra with a continuum model,  \verb|tbabs*thcomp*bbodyrad| in XSPEC, i.e., a blackbody emission Comptonized by thermal electrons (``ThComp''; \cite{zdziarski20}) and attenuated by interstellar absorption. Here, we assume that the Compton cloud completely covers the emission region for the blackbody radiation, which may correspond to the compact star and inner part of the accretion disk. According to the assumption, we fixed the covering fraction of the thermal electrons to 1 in the ThComp model. We also fixed the redshift to 0. Figure \ref{fig:xifi1_total_thcbb} shows the spectra and best-fit model. The fitting yields significant residuals across the entire bandpass, which is particularly prominent in 6 -- 7\,keV (lower panel in figure \ref{fig:xifi1_total_thcbb}). The $\chi^2$ value is unacceptably larger than the degrees of freedom (d.o.f.), $\chi^2/{\rm d.o.f.} = 3634.2/419$. We then introduce 13 Gaussian functions corresponding to emission lines and an absorption edge to the model to compensate for the negative residuals at around 7\,keV. Figure \ref{fig:xifi1_total_thcbb11g1e} shows the fitting results and tables \ref{tab:xifi1_total_thcbb11g1e_cont} and \ref{tab:xifi1_total_thcbb11g1e_line} list the best-fit parameters and fitting statistics. The model reproduces the spectra reasonably well. The normalization of the blackbody component corresponds to the emission radius of seed photons for Comptonization, $19^{+8}_{-3}$\,km, suggesting that the thermal seed photons originate from the neutron star surface and/or the innermost part of the accretion disk. Hereafter, the 13 emission lines (Gaussian models) are referred to as G1 to G13. 

\subsubsection{Line identification}

We identify the origins of the emission lines, on the basis of their line center-energies according to the atomic database AtomDB (\verb|http://atomdb.org/|). Highly ionized atoms of, O, Ne, Si, S, Ar, and Fe are identified (table \ref{tab:xifi1_total_thcbb11g1e_line}) and an emission line and an absorption edge originating from semi-neutral Fe are also found. For the narrow lines among them, G1, G6, G10, G11, G12, and G13, their identifications are unequivocal: O\,VIII K$\alpha$, Si\,XIV K$\alpha$, Fe\,XXV K$\alpha$, Fe\,XXVI K$\alpha$, Fe\, XXV K$\beta$, and Fe\,XXVI K$\beta$, respectively. By contrast, identifications of broad lines are not straightforward. For G2, G3, G4, G5, and G8, the broadening can be explained with a mixture of emission lines. In the close vicinity of these lines, many lines are present with separations smaller than the energy resolution of the XISs. In particular, L transitions of ionized Fe scatter in the $< 2$\,keV energy band. We list ions that have high emissivities within an energy range of $\sigma_{\rm line}$ from respective center energy $\sim E_{\rm line}$ for  G2, G3, G4, G5, and G8 in table \ref{tab:xifi1_total_thcbb11g1e_line}. As for G7, which is another line appearing to be broad, the broadening is statistically significant with $\sim 3 \sigma$ of significance in the fitting. There are, however, theoretically no other prominent lines than the primary line S\,XVI K$\alpha$ around $E_{\rm line}$. Then, we fit the XISFI and XIS1 spectra independently with the same model and find that the widths $\sigma_{\rm line}$ significantly differ between them, $< 65$\,eV and $91^{+25}_{-19}$\,eV, respectively, whereas the values of $E_{\rm line}$ are consistent between them, $2.62^{+0.01}_{-0.03}$\,keV and $2.61 \pm 0.03$\,keV, respectively. The line center energy is close to one of the M edges of gold (2.74\,keV), which is deposited on the surface of the X-ray telescopes. Given that the calibration uncertainty tends to be large at the energies where the effective area varies greatly for a small difference in energy, like gold edges for $Suzaku$, we suspect that the apparent inconsistency of the line widths between the detectors is caused by the calibration uncertainty. We thus decide not to discuss the line width of G7 in the rest of this paper.

\begin{figure}[htbp]
 \begin{center}
  \includegraphics[width=8cm]{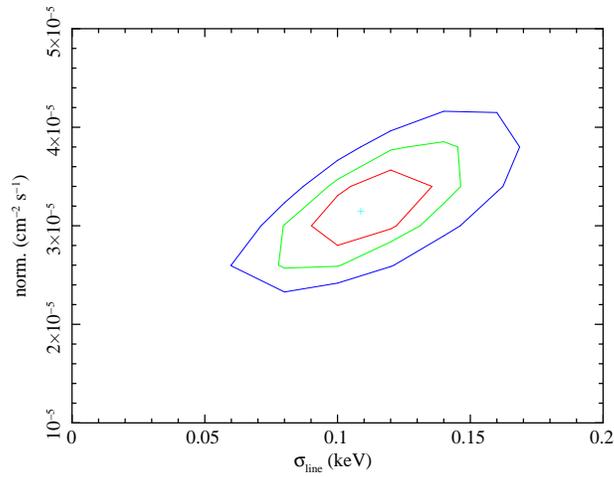} 
 \end{center}
\caption{$\chi^2$ contour plot between the line width $\sigma_{\rm line}$ and the normalization of G9 (semi-neutral Fe\,K$\alpha$) obtained from the fitting of the average spectra. The sky-blue cross denotes the best-fit parameter, whereas the red, green, and blue lines do $\Delta\chi^2 =$ 2.31 ($1\sigma$), 6.17 ($2\sigma$), and 11.8 ($3\sigma$), respectively.}\label{fig:sigma_norm}
\end{figure}

The line G9, which may correspond to semi-neutral Fe K$\alpha$, also shows a width of $0.11 \pm0.02$\,keV. We consider that the broadness is intrinsic to the line from the following reasons. The K-line energy does not significantly vary according to the ionization state of iron for Fe\,I -- XVII and it gradually increases for Fe XVIII and higher ionization degrees, i.e., only very-highly-ionized iron (\cite{kallman04}). Hypothetically, a mixture of highly-ionized iron with less-ionized or neutral iron would appear to be broad. However, such a mixture should result in not only broadening but also a shift of the line center energy to higher energy. In reality, the derived $E_{\rm line}$ of G9 is consistent with that from neutral iron and hence the hypothesis is rejected. Another possibility for the apparent broad width of the line is a slight gain difference between the XISs which could lead to apparent broadening in the summed spectrum. In order to examine the possibility, we make an average spectrum for each FI detector (XIS0 and XIS3) and fitted the XIS0, 1, and 3 spectra simultaneously with the same model but with different constraints for the fitting, that is, the value of $E_{\rm line}$ and normalization of G9 are allowed to vary independently for each detector, while its $\sigma_{\rm line}$ is linked for the data of the three detectors. Table \ref{tab:fe_gain} lists the best-fit results of $E_{\rm line}$ and normalization of G9 for the three instruments. The gain is well aligned among the detectors. The value of $\sigma_{\rm line}$ is obtained to be $0.11 \pm 0.03$\,keV, which indicates that the line is significantly broad. Furthermore, the two adjacent lines to G9, G10 and G11, are both narrow with upper limits of $\sim 30$\,eV in $\sigma_{\rm line}$. Since that the gain is extremely unlikely to vary significantly in such a narrow range of 6.4 -- 7.0\,keV for G9, G10, and G11. The hypothesis of potential gain mismatch at the energy of G9 is rejected. In addition, we show a $\chi^2$ contour plot with respect to $\sigma_{\rm line}$ and normalization of G9 in figure \ref{fig:sigma_norm}. The plot shows that a narrow line ($\sigma_{\rm line} = 0$) is rejected with more than $3\sigma$ of confidence level. We conclude that the semi-neutral Fe K$\alpha$ line is intrinsically broad and that the broad width is not due to line blending. The line width $\sigma_{\rm line} = $0.09 -- 0.11\,keV in the 90\% confidence range corresponds to the line-of-sight velocity dispersion of 4200 -- 5100\,km\,s$^{-1}$, providing that it is caused by Doppler broadening.

\section{Discussion}

\subsection{Comparison with previous works}
In this work, thirteen emission lines were detected in the time-average spectrum. Here, we compare the lines with the results of the previous works, specifically the $Chandra$ and $XMM$ observations conducted by \citet{kallman03}. In their $Chandra$/HETG spectral analysis, eight emission lines were found in the 1 -- 7\,keV band at 1.02\,keV, 1.17\,keV, 1.45\,keV, 2.01\,keV, 2.62\,keV, 6.40\,keV, 6.65\,keV, and 6.90\,keV, which correspond to G3, G4, G5, G6, G7, G9, G10, and G11, respectively, in our result. Their line identifications are also consistent with our result, except the 1.17\,keV line (G4; 1.16\,keV), which they identified as Ne\,X K$\beta$. The theoretical energy of Ne\,X K$\beta$ line is 1.2110 -- 1.2112\,keV according to the AtomDB. This is inconsistent both with our result and with \citet{kallman03} itself. While we identified G4 as L-complex of ionized iron in section 2.2.2, it is worth considering the possibility that G4 is not a line, but an absorption edge. Replacing the model G4 with an edge results in a similar quality of the fit with $\chi^2/{\rm d.o.f} = 852.48/386$. The best fit parameters of the edge are $1.22 \pm 0.02$\,keV for $E_{\rm edge}$ and $0.05 \pm 0.02$ for absorption depth. The energy is consistent with Ne\,XI K edge (1196\,eV) within the statistical uncertainty. However, if the Ne edge is seen, O\,VII K edge should also be seen because oxygen is more abundant than neon in general. The limited energy resolution of XISs does not allow us an unambiguous determination of the feature.

\citet{kallman03} also reported the result of RGS spectral analysis: the three lines detected in the 1.0 -- 1.5\,keV band are consistent with our result, as well as their HETG result, corresponding to G3, G4, and G5 in this work. Their O\,VIII K$\alpha$ line detected at 0.652\,keV is consistent with G1 in our result. The lines detected woth the RGS in the 0.5 -- 1.0\,keV band except O\,VIII K$\alpha$ are not found in this work, probably due to a lower energy resolution of the $Suzaku$ XISs. In their $XMM$/EPIC-pn analysis, five lines were detected in the 6 -- 9\,keV band. The detected lines are consistent with our result, G9 to G13. They also reported the significant width of the neutral-like Fe K$\alpha$ to be $0.29 \pm 0.15$\,keV, for which the confidence level of the error was not stated explicitly. The line width is slightly larger than our current result. Possible time variations of the X-ray emitting region may be the cause of the difference. The lines G2 at 0.885\,keV (Ne\,XI K$\alpha$) and G8 at 3.1\,keV (possibly Ar\,XVII K$\alpha$ and/or S\,XVI K$\beta$) were not reported in their result. The difference could also be due to times variation of the emission region.

Various emission lines we detected may be produced in the photoionized plasma. Because our results of spectroscopic analysis are basically consistent with those of \citet{kallman03}, we can adopt their suggestion for the origin of the emission lines from highly-ionized ion. They applied two-component XSTAR model to estimate the property of the highly-ionized gas, which results in the photoionzation parameter of log$\xi = 1$ and $4.5$ (See \cite{kallman03} for detail).

\subsection{Origin of the neutral-like Fe K$\alpha$ line and geometry of the Compton cloud}

\begin{figure}[htbp]
 \begin{center}
  \includegraphics[width=8cm]{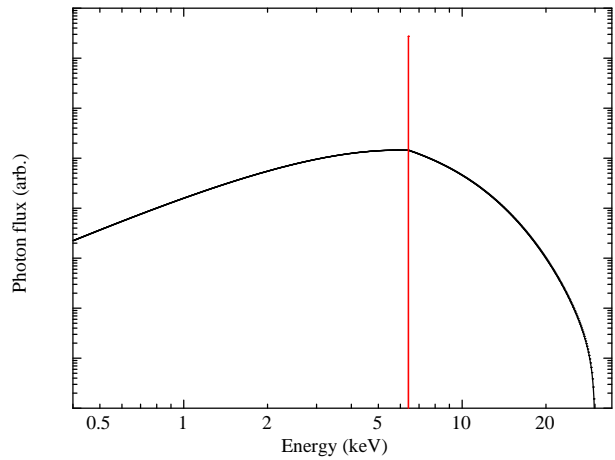} 
 \end{center}
\caption{Simulated spectrum ($in$ $black$) of the reprocessed neutral Fe K$\alpha$ line emission ($in$ $red$) that is scattered in the gas with $kT_{\rm e} = 2$\,keV and $\tau = 7$ and is reprocessed into a continuous emission.}\label{fig:thcomp_feka}
\end{figure}

The emission region of semi-neutral iron should have a low temperature and high density such that thermal ionization and photoionization are supressed. For this reson, the most probable site of origin may be the accretion disk and not the Compton cloud. According to our spectral fitting results, the Compton cloud has temperature of 2\,keV, which is too high to keep iron in neutral or low-ionization-degree state. When the inner part of the accretion disk is covered with the Compton cloud, it may emit the semi-neutral iron line through irradiation from the Compton cloud. However, in this case, the line emission should be scattered by the Compton cloud before it reaches us. An optical depth $\tau$ of the Compton cloud is estimated with an equation

\begin{equation}
\Gamma + \frac{1}{2} = \left[ \frac{9}{4} + \frac{1}{ (kT_{\rm e}/m_{\rm e}c^2) \tau (1 + \tau/3) } \right]^{1/2},
\end{equation}

\noindent as formulated by \citet{zdziarski96}. Substituting the best-fit values of the average spectrum to the equation yields $\tau \sim 21$. Such optically-thick gas for electron scattering will reprocess the line emission and broaden it to a few keV of width. Figure \ref{fig:thcomp_feka} shows the simulated spectrum of the reprocessed 6.4\,keV emission line in a gas with $kT_{\rm e} = 2$\,keV and $\tau = 21$, obtained with the ThComp model in XSPEC. This is greatly broader than and is very different from the observed iron line. Therefore, the semi-neutral iron line should not originate in the inner part of the accretion disk covered by the Compton cloud. Although the optical depth toward the outside of the ADC may be smaller than $\tau = 21$, depending on the emitting site in the accretion disk, the total optical depth $\tau = 21$ is so large that almost none of the regions covered by the Compton cloud can produce the small width of the semi-neutral iron line as we have detected. Considering the solid angle subtended by the Compton cloud, a part of the accretion disk immediately outside the Compton cloud is most likely to be responsible for the semi-neutral Fe\,K$\alpha$ line at 6.4\,keV. The broadening of the line is then interpreted as a Keplerian motion of the accretion disk at the immediate outside of the ADC, i.e., the Keplerian radius may be the rough estimation for the size of the Compton cloud. The Doppler velocity dispersion of the line of 4200 -- 5100\,km s$^{-1}$ corresponds to $(0.8$ -- $1.2) \times 10^9$\,cm of orbital radius for the compact star mass of $1.44\,M_{\rm \odot}$. 

Next, we consider the ionization state of the plasma located at the distance estimated above. Since the photoionization is the main process to determine the ionization state, we estimate the photoionization parameter

\begin{equation}
\xi = \frac{L_{\rm ion}}{n R^2} ,
\end{equation}

\noindent where $L_{\rm ion}$ is the ionizing X-ray luminosity, $n$ is the number density of the irradiated gasses, and $R$ is the distance from the X-ray source. The Fe K-line energy is almost constant for Fe\,I -- XVII, which is consistent with the line energy of the semi-neutral Fe\,K$\alpha$ in our result $E_{\rm line} = 6.40\pm0.02$\,keV. We then consider that the iron atoms in the accretion disk at $\sim 10^9$\,cm should have an ionization state(s) lower than the 17th. According to \citet{kallman04}, the condition is satisfied when $\xi < 100$. We calculate the ionizing X-ray luminosity with the time-averaged spectrum in 7.1 -- 15.0\,keV energy band. Since the X-ray flux decreases rapidly toward the higher energy in the energy-spectrum spece, the ionizing flux is insensitive to the selection of the upper bound. We then obtain a lower limit of the gas density to be $n > 3.1 \times 10^{14}$\,cm$^{-3}$ with $L_{\rm ion} = 1.8 \times 10^{35}$\,erg s$^{-1}$ (where a distance of $D = 7$\,kpc is assumed) and $R = 2.4 \times 10^9$\,cm. In the LMXB system, this degree of high density is present only in the accretion disk. If we assume a standard accretion disk, physical parameters of the accretion disk at the distance $r$ from the compact star are determined with the compact star mass $M$ and an accretion rate $\dot{M}$. The number density of the standard accretion disk is given by 

\begin{equation}
n = 10^{25} \alpha^{-7/10} \left( \frac{M}{M_\odot} \right)^{-7/10} \left( \frac{\dot{M} c^2 }{L_{\rm Edd}} \right)^{11/20} \left( \frac{r}{r_{\rm s}} \right)^{-15/8} {\rm cm^{-3}},
\end{equation}

\noindent where $\alpha$ is a viscosity parameter, $r_{\rm s}$ is the Schwarzschild radius, and $L_{\rm Edd}$ is the Eddington luminosity (\cite{shakura73}). Here, $\dot{M}$ is estimated with the approximation $L_{X} \sim GM\dot{M}/2R_{\rm NS}$. In this case, we obtain $\dot{M} = 4 \times 10^{15}$\,g cm$^{-3}$ for $L_{\rm X} = 4 \times 10^{35}$\,erg s$^{-1}$ and $R_{\rm NS} = 12$\,km (see section 2). Assuming $\alpha = 0.1$ (e.g., \cite{kotko12}) and $L_{\rm Edd} = 10^{38}$\,erg s$^{-1}$, we obtain $n \sim 4 \times 10^{18}$\,cm$^{-3}$ at $10^9$\,cm ($\sim 2\times10^3 r_{\rm s}$) for $M = 1.44 M_\odot$. The derived density is much higher than the lower limit of the number density obtained with the photoionization parameter. In this circumstance, the accretion disk remains well in the low ionization state. Similarly, the temperature of the standard accretion disk is given by

\begin{equation}
kT = 6 \alpha^{-1/5}  \left( \frac{M_{\rm 1}}{M_\odot} \right)^{-1/5} \left( \frac{\dot{M} c^2 }{L_{\rm Edd}} \right)^{3/10} \left( \frac{r}{r_{\rm s}} \right)^{-3/4} {\rm keV}.
\end{equation}

\begin{figure}[htbp]
 \begin{center}
  \includegraphics[width=8cm]{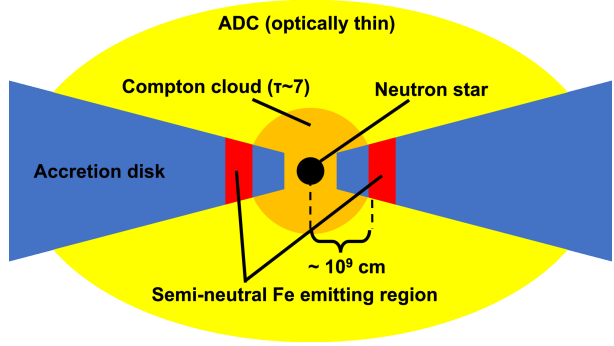} 
 \end{center}
\caption{Schematic view of 2S\,0921$-$630. Note that the ellipsoid shape of the ADC is tentative one. Its actual shape is not constrained.}\label{fig:scem}
\end{figure}

\noindent The temperature at $2\times 10^3 r_{\rm s}$ is then calculated to be 0.01\,keV. This is cold enough for iron atoms not to be ionized by the thermal energy. We show a schematic view of the system suggested discussed in this section in figure \ref{fig:scem}.

\section{Summary}

We analyzed $Suzaku$ archive data of the ADC source 2S\,0921$-$630. The average spectrum was reproduced with a blackbody emission with $kT = 0.3$\,keV Comptonized with an optically-thick gas (Compton cloud) with $kT_{\rm e} = 2$\,keV and $\tau \sim 21$, combined with thirteen emission lines, and an absorption edge. Twelve out of the thirteen emission lines were identified as highly ionized atoms, O, Ne, Si, S, Ar, and Fe, whereas the remaining one was consistent with neutral-like Fe K$\alpha$ with a significant width of $90$ -- $110$\,eV (90\% confidence range in $\sigma$), corresponding to an orbital radius of $(0.8$ -- $1.2) \times 10^9$\,cm. Most of the detected lines are consistent with the previous work by \citet{kallman03}, whereas Ne\,IX K$\alpha$ and possible Ar\,XVII K$\alpha$ and/or S\,XVI K$\beta$ lines that we detected were not reported by them. We suggested that the semi-neutral iron K$\alpha$ line originated from the accretion disk in the immediate outside of the Compton cloud, i.e., we constrained the size of the Compton cloud to be ($0.8$ -- $1.2$) $\times 10^9$\,cm. The absorption edge was also identified as that from the semi-neutral iron.



\begin{ack}
This work is supported by Japan Society for the Promotion of Science (JSPS) KAKENHI Grant Numbers JP21K20372 and JP19K03904.
\end{ack}





\bibliographystyle{apj}
\bibliography{main}




\newpage


\newpage



\begin{table}[htbp]
  \tbl{$Suzaku$ observations of 2S\,0921$-$630}{%
  \begin{tabular}{ccccc}
\hline														
OBSID	&	Observation start	&	Observation end	&	Exposure [s]	&	Orbital phase$^*$ \\	\hline
402059010	&	2007-08-23 21:17	&	2007-08-24 23:50	&	36499	&	0.32 -- 0.44 \\
402060010	&	2007-08-26 01:13	&	2007-08-27 03:50	&	40350 	&	0.56 -- 0.68 \\
402057010	&	2007-08-28	18:16	&	2007-08-29 20:00	&	35556	&	0.86 -- 0.98 \\
402058010	&	2007-08-30 09:47	&	2007-08-31 12:19	&	39499	&	0.04 -- 0.16 \\ \hline
 \end{tabular}}\label{tab:obs}
\begin{tabnote}
$^*$Derived with the ephemeris provided by \citet{ashcraft12}.
\end{tabnote}
\end{table}

\begin{table}[htbp]
  \renewcommand{\thetable}{\arabic{table}a}
  \tbl{Best-fit parameters of the average spectrum: continuum components and statistics}{
  \begin{tabular}{cccc}
\hline
Model component	&	Parameter	&	Value	&	Note \\ \hline
TBabs	&	$N_{\rm H}$ [$10^{22}$\,cm$^{-2}$]	&	$0.091^{+0.003}_{-0.004}$	&	\\
ThComp	&	$\Gamma$	&	$1.443^{+0.003}_{-0.004}$	&	\\
	&	$kT_{\rm e}$	&	$1.99 \pm 0.02$	&	\\
	&	Covering fraction	&	1	&	Fixed	\\
	&	Redshift	&	0	&	Fixed	\\
bbodyrad	&	$kT_{\rm bb}$ [keV]	&	$0.23^{+0.02}_{-0.04}$	&	\\
	&	norm. [km$^2$ at 10\,kpc]	&	$(7^{+8}_{-2}) \times 10^2$	&	 	\\ \hline
	&	$F_{\rm X}^*$	[erg s$^{-1}$ cm$^{-2}$]	&	$(8 \pm 1) \times 10^{-11}$	&	\\
	 &	$L_{\rm X}^\dagger$	[erg s$^{-1}$]	&	$(4.7 \pm 0.8) \times 10^{35}$	&	\\ \hline
	&	$\chi^2$&	851.32	&	\\
	&	d.o.f.	&	385	&	\\ \hline
  \end{tabular}}\label{tab:xifi1_total_thcbb11g1e_cont}	
\begin{tabnote}
$^*$Total X-ray flux calculated in the 0.4--10.0\,keV energy range, $^\dagger$total X-ray luminosity calculated for $F_{\rm X}$ and $D = 7$\,kpc.
\end{tabnote}
\end{table}

\begin{table}[htbp]
  \addtocounter{table}{-1}
  \renewcommand{\thetable}{\arabic{table}b}
  \tbl{Best-fit parameters of the average spectrum: emission lines and an absorption edge}{
  \begin{tabular}{cccccc}
\hline
Model component	&	$E_{\rm line, edge}$	 [keV]	&	$\sigma_{\rm line}$ [keV]		&	norm./optical depth$^*$	&	Equivalent width [eV]	&	ID \\ \hline
Gaussian1 (G1)	&	$0.661^{+0.006}_{-0.008}$	&	$< 0.02$	&	$7 \pm 2$		&	$7 \pm 2$	&	O\,VIII K$\alpha$	\\ 
Gaussian2 (G2)	&	$0.885 \pm 0.005$	&	$0.023^{+0.007}_{-0.013}$	&	$9 \pm 2$	&	$11^{+5}_{-3}$	&	 Ne\,IX K$\alpha$ + Fe\,XVII -- XIX L	\\
Gaussian3 (G3)	&	$1.02 \pm 0.01$	&	$0.034 \pm 0.006$	&	$10 \pm 2$		&	$13 ^{+5}_{-2}$	&	Ne\,X K$\alpha$ + Fe\,XVII -- XXIII L	\\ 
Gaussian4 (G4)	&	$1.16 \pm 0.01$	&	$< 0.02$	&	$1.5 \pm 0.5$		&	$2 \pm 1$	&	Fe\,XVII -- XXIV L	\\ 
Gaussian5 (G5)	&	$1.42 \pm 0.02$	&	$0.07 \pm 0.01$	&	$4.1^{+1.0}_{-0.6}$		&	$8 ^{+4}_{-2}$	&	Mg\,XII K$\alpha$ + Fe\,XXI -- XXIII L	\\
Gaussian6 (G6)	&	$2.000^{+0.009}_{-0.003}$	&	$< 0.01$	&	$2.8 \pm 0.4$		&	$8^{+2}_{-1}$	&	Si\,XIV K$\alpha$	\\
Gaussian7 (G7)	&	$2.60 \pm 0.02$	&	$0.08 \pm 0.01$	&	$4.2 \pm 0.5$		&	$16^{+4}_{-2}$	&	S\,XVI K$\alpha$	\\
Gaussian8 (G8)	&	$3.10 \pm 0.02$	&	$< 0.05$	&	$1.2 \pm 0.3$		&	$5^{+2}_{-1}$	&	Ar\,XVII K$\alpha$ + S\,XVI K$\beta$	\\
Gaussian9 (G9)	&	$6.40 \pm 0.02$	&	$0.11 \pm 0.02$	&	$3.1 \pm 0.3$		&	$37^{+11}_{-8}$	&	Fe\,I -- XVII K$\alpha$ (semi-neutral)	\\
Gaussian10 (G10)	&	$6.692 \pm 0.004$	&	$< 0.03$	&	$5.3 \pm 0.3$		&	$67^{+14}_{-3}$	&	 Fe XXV K$\alpha$	\\
Gaussian11 (G11)	&	$6.968^{+0.003}_{-0.005}$	&	$< 0.01$		&	$4.3 \pm 0.1$		&	$62^{+20}_{-12}$	&	 Fe\,XXVI K$\alpha$	\\
Gaussian12 (G12)	&	$7.83^{+0.03}_{-0.02}$	&	$< 0.05$		&	$1.2 \pm 0.3$		&	$25^{+10}_{-6}$	&	Fe\,XXV K$\beta$	\\
Gaussian13 (G13)	&	$8.27^{+0.02}_{-0.01}$	&	$< 0.1$		&	$0.8 \pm 0.3$		&	$18^{+10}_{-6}$	&	 Fe\,XXVI K$\beta$	\\
Edge	&	$7.1 \pm 0.1$	&	-	&	$0.08^{+0.01}_{-0.02}$	&	-	&	Fe\,I -- IV K edge (semi-neutral)	\\ \hline
  \end{tabular}}\label{tab:xifi1_total_thcbb11g1e_line}	
\begin{tabnote}
$^*$Normalization in units of $10^{-5}$\,photons\,cm$^{-2}$\,s$^{-1}$ for Gaussians, absorption depth at $E_{\rm line}$ for Edge.
\end{tabnote}
\end{table}

\begin{table}[htbp]
  \tbl{Best-fit parameters of G9 for each instrument}{%
  \begin{tabular}{cccc}
\hline														
Instrument	&	$E_{\rm line}$ [keV]	&	$\sigma_{\rm line}$ [keV]	&	norm.$^*$	\\	\hline
XIS0	&	$6.40 \pm 0.03$	&	$0.11 \pm 0.03$	&	$3.6 \pm 0.6$ \\
XIS1	&	$6.39^{+0.04}_{-0.02}$	&	(linked to XIS0)	&	$3.1^{+0.8}_{-0.7}$ \\
XIS3	&	$6.41 \pm 0.03$	&	(linked to XIS0)	&	$3.2^{+0.7}_{-0.6}$ \\ \hline
 \end{tabular}}\label{tab:fe_gain}
\begin{tabnote}
$^*10^{-5}$\,photons\,cm$^{-2}$\,s$^{-1}$.
\end{tabnote}
\end{table}


\end{document}